\documentclass[11pt,a4paper]{article}

\usepackage{amsmath,amssymb}
\usepackage{epsfig,graphicx}
\usepackage{subfigure}
\usepackage{graphicx}
\usepackage{rotating}
\usepackage{cancel}
\usepackage{bm}
\usepackage{color}
\usepackage{comment}
\usepackage{cite}
\usepackage{psfrag}

\newcommand{\nocc}{{{f}}}
\newcommand{\ndens}{{n}}
\newcommand{\Qs}{{Q}}
\newcommand{\fax}{{F}}
\newcommand{\dd}{\mathcal{D}}
\newcommand{\ddzero}{\dd_{\rm ini}}
\newcommand{\ns}{n}
\newcommand{\nocczero}{{{\nocc_{\rm ini}}}}
\newcommand{\nocctyp}{{{\nocc_{\rm typ}}}}
\newcommand{\noccfudge}{{{\nu}}}
\newcommand{\tcond}{t_{\rm cond}}
\newcommand{\ttherm}{t_{\rm therm}}
\newcommand{\tini}{t_{\rm ini}}
\newcommand{\aini}{a_{\rm ini}}
\newcommand{\Hini}{H_{\rm ini}}
\newcommand{\ptyp}{p_{\rm typ}}

\renewcommand\[{\left[}

\newcommand{\exclude}[1]{}

\def\beq{\begin{equation}}
\def\eeq{\end{equation}}

\begin{document}
\numberwithin{equation}{section}
\title{
\vspace{2.5cm} 
\Large{\textbf{Far from equilibrium dynamics \\of Bose-Einstein condensation for Axion Dark Matter 
\vspace{0.5cm}}}}

\author{J\"urgen Berges, and Joerg Jaeckel\\[2ex]
\small{\em Institut f\"ur Theoretische Physik, Universit\"at Heidelberg, Philosophenweg 16, 69120 Heidelberg, Germany}\\[0.5ex]
}

\date{}
\maketitle

\begin{abstract}
\noindent
Axions and similar very weakly interacting particles are increasingly compelling candidates for the cold dark matter of the universe. Having very low mass and being produced non-thermally in the early Universe, axions feature extremely high occupation numbers.
It has been suggested that this leads to the formation of a Bose-Einstein condensate with potentially significant impact on observation and direct detection experiments. 
In this note we aim to clarify that if Bose-Einstein condensation occurs for light and very weakly interacting dark matter particles, it does not happen in thermal equilibrium but is described by a far-from-equilibrium state.
In particular we point out that the dynamics is characterized by two very different timescales, such that condensation occurs on a much shorter timescale than full thermalization.
\end{abstract}

\newpage

\section{Introduction}
Although dark matter makes up approximately one quarter of the total energy content of the Universe we still know very little about it. It is most likely not Standard Model matter and it interacts with Standard Model particles as well as with itself only very weakly. Moreover, it behaves much like a gas of non-relativistic particles, i.e. particles with very little kinetic energy, hence cold dark matter.

Therefore, it is not surprising that the two perhaps best known (and best motivated) candidates for dark matter particles, axions and weakly interacting massive particles (WIMPs), have very different properties. 
Axions are very light, sub-eV bosons whereas (SUSY) WIMPs are heavy, multi-GeV fermions. Axions are produced non-thermally from coherent oscillations of the axion field~\cite{Preskill:1982cy,Abbott:1982af,Dine:1982ah} far from thermal equilibrium, whereas WIMPs are produced in thermal collisions in or close to thermal equilibrium.
Nevertheless, both are thought to behave in many ways like a gas of cold, non-relativistic particles.

However, recently it has been suggested~\cite{Sikivie:2009qn} that there is indeed a property that sets axions apart from WIMPs: 
Axions are bosons and if they are dark matter their low momentum states feature very high occupation numbers. This suggests the possibility
that they form a Bose-Einstein condensate. The resulting collective behavior may lead to distinct observational signatures~\cite{Sikivie:2012gi,Banik:2013rxa} as well as 
having important effects for experiments for the direct detection of axion dark matter~\cite{Duffy:2006aa,Irastorza:2011gs,Jaeckel:2013sqa}. 
Usually Bose-Einstein condensation is treated in thermal equilibrium, and indeed the arguments 
of Refs.~\cite{Sikivie:2009qn,Erken:2011vv,Erken:2012dz,Saikawa:2012uk} are based on estimates for thermalization rates\footnote{Indeed the authors of the more recent papers~\cite{Davidson:2013aba,Noumi:2013zga} more carefully refer to dissipation rates and self interaction rates.}. However, the extremely weak self-interactions of axions suggest that it takes very long times to reach full thermal equilibrium and establish a proper thermal distribution.

In this note we want to clarify this apparent conflict and argue that, while full thermal equilibrium\footnote{We are considering only axions. When talking about full thermal equilibrium we mean thermal equilibrium of the axions by themselves. Our arguments do not concern the question of a possible equilibration with other species such as photons.} is probably not reached, a non-thermal Bose-Einstein condensate may nevertheless be formed. Based on results of non-equilibrium field theory~\cite{Berges:2012us,Micha:2004bv,Nowak:2013juc,BS}, we elucidate the non-equilibrium nature of such a condensate: 
\begin{itemize}
\item{} We clearly distinguish between the very different timescales for condensation $t_{\rm cond}$ and that for thermalization $t_{\rm therm}$
with $\tcond\ll\ttherm$.
\item{} We estimate the time needed to reach a fixed value of the condensate fraction $N_0/N_{\rm total}$ depending on the spatial volume. 
\item{} The non-thermal condensate fraction can differ from that in thermal equilibrium. More importantly we note that the condensate is in a strongly correlated state which significantly deviates from a weakly interacting gas approximated by free field theory.
\end{itemize}

\bigskip

Qualitatively, similar behavior may happen for other very light bosonic dark matter candidates such as axion-like particles (ALPs) or hidden photons~\cite{Arias:2012ly} (see~\cite{Jaeckel:2010ni} for a review of such particles).  For the purposes of this work the main feature of axions is that they are a very light bosonic particles with extremely weak self-interactions (we mostly neglect the similarly weak interactions with other matter particles). Therefore, our treatment should apply directly to axions and axion-like particles but similar mechanisms may apply for more general very light bosonic dark matter particles as well. For simplicity we will nevertheless refer to our particles as ALPs.
 
In the following we will only consider ALPs interacting via their self-interactions.
As already found in~\cite{Sikivie:2009qn} these interactions are probably not sufficient to achieve Bose-Einstein condensation of ALPs in the early Universe.
The phenomenologically more relevant interactions are probably those via gravitons~\cite{Sikivie:2009qn,Erken:2011vv,Erken:2012dz,Saikawa:2012uk,Davidson:2013aba,Noumi:2013zga}.
Therefore our toy model does not allow to address the quantitative question whether the gravitational interactions of ALPs are fast enough to achieve condensation (both positive~\cite{Sikivie:2009qn,Erken:2011vv,Erken:2012dz,Saikawa:2012uk,Noumi:2013zga} 
and negative\cite{Davidson:2013aba} evidence exists).
Importantly however, we expect that our general arguments hold and the far from equilibrium nature with the features described above will persist if such a Bose-Einstein condensate is formed via gravitational interactions.

\section{ALP dark matter}
We briefly recall how very light dark matter particles can be a good dark matter candidate.
The simplest production mechanism for ALP dark matter is the so-called misalignment mechanism~\cite{Preskill:1982cy,Abbott:1982af,Dine:1982ah}.
The essential features of this mechanism can be directly understood by looking at the evolution of a free scalar field in the expanding Universe. 
The basic principle and the fact that axions produced in this manner behave like dark matter can easily be understood by looking at the evolution of a constant field. We will return to the issue of initial fluctuations in the next Sect.~\ref{highocc}. The equation of motion for the field $\phi(t)$ is then,
\begin{equation}
\ddot{\phi}+3H\dot{\phi}+m^{2}(t)\phi=0,
\end{equation}
where $H$ is the Hubble constant and the dot denotes a time derivative.
This is the equation of a damped harmonic oscillator. We have allowed for a time-dependent mass $m(t)$ to explicitly include the case of the axion.

At early times where $H\gg m$ the oscillator is overdamped and the field is essentially stuck at its initial value. This also justifies why we start with a non-vanishing initial value: the field simply did not have time to relax to its minimum value.
At a later time $t_{1}$, when $3H(t_{1})\approx m(t_{1})$ the field begins to oscillate.
In a WKB-like approximation the solution then is
\begin{equation}
\label{latesol}
\phi(t)=\phi_{1}\left(\frac{m_{1}a^{3}_{1}}{m(t) a^{3}(t)}\right)^{1/2}\cos\left(\int^{t}_{t_{1}}dt \,m(t)dt\right).
\end{equation}
Here $a$ denotes the scale factor and the index $1$ indicates that a quantity is evaluated at time $t_{1}$.  

Inserting Eq.~\eqref{latesol} into the expression for the energy density of a scalar field we find,
\begin{equation}
\rho(t)=\frac{1}{2}\left(\dot{\phi}^2+m^2(t)\phi^2\right)=\frac{1}{2}\phi^{2}_{1}m_{1}m(t)\left(\frac{a_{1}}{a(t)}\right)^{3}
=\rho_{1}\left(\frac{m(t)}{m_{1}}\right)\left(\frac{a_{1}}{a(t)}\right)^{3}.
\end{equation}
For constant $m(t)\approx const$ the energy density is diluted with the increasing volume. This is exactly what we expect for non-relativistic matter, i.e. dark matter.
Indeed one can check that the number density is
\begin{equation}
n(t)=\frac{\rho(t)}{m(t)}=n_{1}\left(\frac{a_{1}}{a(t)}\right)^{3},
\end{equation} 
as we would expect from adiabatic expansion.

For the simplest case of ALPs, the mass $m$ will indeed be constant. For the QCD axion, whose mass arises during the QCD phase transition, the mass is 
time dependent during this period, which has some effect on the final density. However, at much later times, during the release of the cosmic microwave background and then during structure formation the mass of the QCD axion is also constant and no changes in these observations are expected from this effect.

To be concrete, for the axion one finds
an axion population whose contribution, in terms of the critical density is~\cite{Sikivie:2006ni}
\begin{equation}
\Omega_{a,0}h^2=\kappa_{a}\left(\frac{\fax_{a}}{10^{12}\,{\rm GeV}}\right)^{1.175}\theta^{2}_{i},
\end{equation}
where $\fax_{a}$ is the axion decay constant, $0.5\lesssim \kappa_{a}\lesssim {\rm few}$ and $\theta_{i}$ is the initial misalignment angle $\theta_{i}\in [0,2\pi]$.
The index $0$ indicates that a quantity is evaluated today.
To avoid having to fine-tune $\theta_{i}$ to a small value suggests ${\mathcal O}(\fax_{a})\sim 10^{12}\,{\rm GeV}$.

The axion mass today is given by
\begin{equation}
m_{a,0}\approx 6\,\mu{\rm eV}\left(\frac{10^{12}\,{\rm GeV}}{\fax_{a}}\right).
\end{equation}
Assuming that all of today's dark matter is made of ALPs, fixes the average ALP density to be
\begin{equation}
\rho_{\rm 0}\approx 1.2\frac{{\rm keV}}{{\rm cm}^{3}}.
\end{equation}
We note that this is the global average density and not the local density in our galaxy.

\section{High occupation numbers, interaction rates and the question of thermal equilibrium}\label{highocc}
We now turn to an estimate of the occupation numbers for axion dark matter (see also~\cite{Sikivie:2009qn}). The huge number that we will find then motivates a closer look at the Bose enhanced interaction rates and the question of thermal equilibrium.

\subsection{Occupation numbers for ALP dark matter}
To estimate the occupation number we also need information about the momentum distribution.
For concreteness we discuss the situation for axions. Here we need to distinguish two scenarios depending on whether the Peccei-Quinn phase transition happened before or after inflation (see, e.g.~\cite{Sikivie:2006ni}).

In the former case the axion field is present during inflation. All fluctuations that are initially present are smoothed out to scales much larger than the horizon. The field is therefore essentially all concentrated in the zero momentum mode and is essentially classical. The only fluctuations present are those $\sim H_{I}$ imprinted by inflation itself. 

In the opposite scenario where the Peccei-Quinn phase transition happens after inflation several different populations can contribute~\cite{Sikivie:2006ni}: the vacuum re-alignment contribution (see previous section) of axions produced by string decay and those from domain wall decay.
The first two populations, which in the following we will use as a guideline for our estimates\footnote{We emphasize that the details of the intial condition are not expected to affect the evolution near the non-thermal fixed point which is discussed below~\cite{Berges:2013lsa}.}, are expected to have typical momenta of the same order of magnitude,
\begin{equation}
\label{momestimate}
\delta p_{1}\sim H_{1}\sim\frac{1}{t_{1}},
\end{equation}
where the index $1$ indicates that a quantity is evaluated at the time $t_{1}$ when the axion field starts to oscillate.
At a later time these initial momenta are reduced by the cosmological expansion such that we have,
\begin{equation}
\label{momentum}
\delta p(t)\sim H_{1}\frac{a(t_{1})}{a(t)}.
\end{equation}

Finally, for completeness we note that axions generated by the decay of domain walls are expected to have a momentum distribution which is broader by a factor of $10^3-10^4$. But numerical estimates suggest that this population is sub-dominant. 
For this reason, but mainly for simplicity and concreteness we will from now on simply use the estimate~\eqref{momestimate} and use it also for the general ALP case.

\bigskip

To obtain an estimate for the typical occupation number $\nocctyp$ we can now simply equally occupy all available states with momenta $\leq \delta p(t)$,
\begin{equation}
\label{occ}
\nocctyp\sim \frac{(2\pi)^3 n(t)}{\frac{4}{3}\pi \delta p^{3}(t)}\sim \frac{6\pi^2n_{0}}{H^{3}_{1}}\left(\frac{a_{0}}{a_{1}}\right)^3
\sim \frac{162\pi^2 n_{1}}{m^{3}_{1}}.
\end{equation}
We note that this is indeed independent of time. As the universe expands the density decreases, but so does the momentum such that the occupation numbers remain the same. This is exactly what we expect for adiabatic expansion.

Before we continue it should be remarked that the above calculation strictly speaking only makes sense somewhat later than the time $t_{1}$ when oscillations start. The reason is that for our estimate we have used the continuum phase space element which only makes sense if the volume is large enough
to permit a large number of states with momentum $\lesssim \delta p(t)$,
\begin{equation}
\frac{4}{3}\pi \frac{(\delta p(t))^3}{(2\pi)^3} V\sim \frac{4}{3}\pi \frac{1}{(2\pi)^3}H^{3}_{1}\left(\frac{a_{1}}{a(t)}\right)^3\frac{4}{3}\pi \frac{1}{H^{3}(t)}
\sim 0.07 \left(\frac{a(t)}{a_{1}}\right)^{3}\gg 1.
\end{equation}
In other words we should have 
\begin{equation}
 \left(\frac{a(t)}{a_{1}}\right)\gtrsim 10.
\end{equation}
However, in the following we will mostly ignore this issue.

Inserting $n_{1}\sim (1/2) m_{1}\phi^2_{1}$ and using $3H_{1}\approx m_{1}$ we find,
\begin{equation}
\label{nocctyp}
\nocctyp=\noccfudge \,\,10^3 \left(\frac{\phi_{1}}{m_{1}}\right)^2 =\noccfudge\,\,  10^{57} \left(\frac{\phi_{1}}{10^{12}\,{\rm GeV}}\right)^2
\left(\frac{\mu{\rm eV}}{m_{1}}\right)^2.
\end{equation}
In light of the very approximate nature of the above estimate here we keep an arbitrary factor of 
$\noccfudge$. The above calculation gives $\noccfudge\sim 1$.  This factor then also allows to account for possible mechanisms that enhance the occupation somewhat compared to the naive estimate, e.g. by reducing the size of the typical initial momentum below $H_{1}$, or by a later evolution of the mass of the particle.

For the very light ALPs, which we are interested in, we always have $\phi_{1}\gg m_{1}$ and therefore very large occupation numbers. This enormous occupation number is now available to compensate extremely small couplings via a Bose enhancement.

\subsection{Self-interaction rates and being far from equilibrium}
To see if the compensation is sufficient we need to estimate the size of the self-interactions. The Lagrangian for our scalar field is given by,
\begin{equation}
{\mathcal L}=\frac{1}{2}(\partial_{\mu}\phi)^{2}-V(\phi).
\end{equation}
Axions and axion-like particles are usually pseudo-Goldstone fields. 
We therefore expect a potential of the form,
\begin{equation}
\label{cospotential}
V(\phi)=m^{2}\fax^{2}\left[1-\cos\left(\frac{\phi}{\fax}\right)\right],
\end{equation}
where $\fax$ is the scale of spontaneous symmetry breaking of the symmetry in question. For true QCD axions this simple form is not quite the full result, because of mixing effects with the pions. Nevertheless it gives the
essential scaling behavior.

Expanding about the minimum we have,
\begin{equation}
V(\phi)=\frac{m^{2}}{2}\phi^2-\frac{1}{4!}\frac{m^{2}}{\fax^{2}}\phi^4+\ldots.
\end{equation}
For our current qualitative discussion the negative sign is not relevant (the potential is also stabilized by higher order interactions). In other words we expect a self-coupling of order,
\begin{equation}
\label{couptyp}
\lambda=\kappa\, \frac{m^2}{\fax^2}=\kappa \,\,10^{-54} \left(\frac{m}{\mu{\rm eV}}\right)^{2}\left(\frac{10^{12}\,{\rm GeV}}{\fax}\right)^{2},
\end{equation}
where $\kappa=1$ for the simple cosine potential discussed above. However, to be general we keep a proportionality factor to allow for an enhanced or reduced coupling. Nevertheless the naive expectation is $\kappa\sim 1$.

To get an estimate\footnote{Strictly speaking this estimate is valid only in the particle kinetic regime, i.e. when the energy dispersion $\sim \frac{1}{2}m\delta v^2$ is larger than the relevant rates. This is not strictly true for all rates of interest in this case. Nevertheless the simple argument we give here demonstrates the relevant point.}  for the interaction rate we can consider two-to-two scattering which for single particles has a cross section
\begin{equation}
\sigma_{0}=\frac{\lambda^2}{64\pi^2 m^2}.
\end{equation}
We first discuss the favorable situation, where we can consider the scattering between two highly occupied states, which is then enhanced by two factors of $\nocctyp$. The relaxation rate {\emph{for such states}} is the scattering rate per particle in these states and is enhanced by a single factor of $\nocctyp$,
\begin{equation}
\label{naiverelax}
\Gamma_{\rm relax, high}\sim n v_{\rm typ} \nocctyp\sigma_{0}.
\end{equation}
The typical velocity $v_{\rm typ}$ can be estimated from the momentum spread Eq.~\eqref{momentum},
\begin{equation}
v_{\rm typ}\sim \frac{\delta p(t)}{m(t)}.
\end{equation}

The relaxation rate now has to be compared to the Hubble scale,
\begin{equation}
\frac{\Gamma_{\rm relax, high}}{H}\sim 0.25\,\noccfudge \,\lambda^2 \,\frac{\phi_{1}}{m_{1}}\,\left(\frac{\phi_{1}}{m(t)}\right)^{3}
\left(\frac{a_{1}}{a(t)}\right)^{2}.
\end{equation}

We now consider this rate for an ALP with constant mass. In this case typical initial values are $\phi_{1}\sim \fax$. 
We then find in agreement with~\cite{Sikivie:2009qn},
\begin{equation}
\label{relaxhigh}
\frac{\Gamma_{\rm relax, high}}{H}\sim 0.25\,\noccfudge\,\kappa^2\left(\frac{a_{1}}{a(t)}\right)^{2}.
\end{equation}
So for $\kappa\sim\nu\sim 1$ and close to the time $t_{1}$ when the $\phi$ starts to behave like dark matter, the relaxation rate is just about the order of the Hubble scale 
and one may expect that this can lead to a significant change in the momentum distribution. We also stress that the above is just a rough estimate and the true rate may be higher or lower by a couple of orders of magnitude. 
However, as the Universe expands the rate drops faster than the Hubble scale and is soon insufficient to cause effective changes.

We now come to an important point concerning the question of thermal equilibrium. The above relaxation rate is for scattering between two highly occupied states with occupation numbers $\sim \nocctyp$.
On the other hand, typical momenta for scattering near thermal equilibrium have an occupancy of order one, since
\begin{equation}
\nocc(p)=\frac{1}{\exp((\omega(p)-\mu)/T)-1}\lesssim 1,\quad{\rm for}\quad \exp((\omega(p)-\mu)\gtrsim 2,
\end{equation}
where the condition on the right hand side indicates the beginning of the exponential tail.
Establishing a proper equilibrium distribution could therefore take a forbiddingly long time, as there is little or no Bose-enhancement, which in the case of interest makes a difference of more than 50 orders of magnitude,
\begin{equation}
\frac{\Gamma_{\rm relax, tail}}{H}\sim \frac{n v_{\rm typ}\sigma_{0}}{H}\lesssim \frac{\kappa^2}\nocctyp \left(\frac{a_{1}}{a(t)}\right)^{2}\ll 1.
\end{equation}
Therefore, we should be prepared to face a far-from-equilibrium situation. In this case quite different processes apart from relaxation, such as non-equilibrium instabilities or turbulence, can play an important role. 

\subsection{The timescale for thermalization}
The above estimates are very crude and only serve to demonstrate qualitatively the fact that one has to deal with far-from-equilibrium dynamics.
In this subsection we determine the parametric dependence of the timescale for thermalization using results from a proper non-equilibrium treatment of such processes.

The non-equilibrium dynamics starting from highly occupied initial states has been studied extensively in recent years \cite{Berges:2012us,Micha:2004bv,Nowak:2013juc,BS}. Characteristic properties of this far from equilibrium dynamics turn out to be universal, such that it is insensitive to the details of the underlying model and initial conditions. Accordingly, very different physical systems on most diverse energy scales ranging from early-universe inflaton dynamics to table-top experiments with cold atoms are described by universal scaling exponents and scaling functions in this case. The universality is based on the existence of non-thermal
fixed points \cite{Berges:2008sr}, which are attractor solutions with self-similar scaling behavior characteristic for wave turbulence. 

In terms of the time-dependent occupation number distribution $\nocc(t,p)$ such a self-similar evolution corresponds to
\begin{equation}
\nocc(t,p) = (Q t)^\alpha \, f_S((Qt)^\beta p) \, ,
\label{eq:scaling}
\end{equation}
where $Q$ denotes some characteristic (initial) momentum scale. The scaling exponents $\alpha$ and $\beta$ as well as the scaling function $f_S$ are universal up to normalizations of the latter. The behavior (\ref{eq:scaling}) represents an enormous reduction of the possible dependence of the dynamics on variations in time and momenta, since it states that $(Q t)^{-\alpha} \nocc(t,p)$ only depends on the product $(Q t)^\beta p$ instead of separately depending on time and momenta. 
The universal distribution function $f_S$ does not change with time. Therefore any characteristic momentum $\ptyp(t)$ then scales as $\ptyp(t) \sim Q (Q t)^{-\beta}$, such that $f_{S}(\ptyp(t)(Qt)^{\beta})=f_{S}(\ptyp(t^{\prime})(Qt^{\prime})^{\beta})$ for any two times $t,t^{\prime}$ in the self-similar regime. 

This scaling behavior can be used to obtain a lower bound on the thermalization time $t_{\rm therm}$ for known exponent $\alpha$. For our purposes we consider a situation at initial time $\tini$ described by some small coupling parameter $\lambda$ with
\begin{equation}
\label{thetainitial}
\nocc(\tini,p)=\nocczero\, \Theta(Q-p) \, \sim \, \frac{1}{\lambda}\, \Theta(Q-p) \, .
\end{equation}  
This choice is motivated by comparing with Eq.~\eqref{naiverelax}. The $1/\lambda$ enhanced occupation number compensates factors of $\lambda$ from the scattering. Moreover, for an ALP with potential as in Eq.~\eqref{cospotential} and an initial field value $\phi\sim \fax$ we obtain from Eq.~\eqref{nocctyp},
\begin{equation}
\nocczero\sim \frac{\rho}{m}\frac{1}{\delta p^3}\sim \frac{m^{2} \fax^2}{m}\frac{1}{m^3}\sim \frac{\fax^2}{m^2}\sim \frac{1}{\lambda},
\end{equation}
up to a typical momentum $Q\sim \delta p\sim H \sim m$. 

Accordingly, initially our distribution exhibits for $\lambda \ll 1$ very large occupancies with typical momentum $\ptyp(\tini)=Q$. Thermalization requires that the occupancy for typical momenta drops to order one as mentioned before, i.e.~$\nocc(t_{\rm therm},\ptyp(t_{\rm therm})) \sim 1$. Using the scaling behavior (\ref{eq:scaling}) to describe the evolution down to order one occupancies one concludes that
\begin{equation}
\label{ttherm}
\ttherm \, \gtrsim \, \frac{1}{Q} \left(\frac{1}{\lambda}\right)^{\frac{1}{|\alpha|}} \, ,
\end{equation}
where we have used $\tini \sim 1/Q$ and that our values for the exponent $\alpha$ are negative. In particular we have for the self-interacting scalar theories $\alpha = -4/5$ for the relativistic case and $\alpha=-3/2$ for the non-relativistic regime~\cite{Micha:2004bv}. 

We conclude that starting from highly occupied initial conditions the timescale for thermalization diverges in the weak coupling limit. It should be emphasized that the reason for the very long time scales for small couplings is directly related to the order-one occupancy of typical momenta in thermal equilibrium. Any Bose condensation process that relies on reaching a thermal distribution starting from such an initial state has to face these long times. However, we will argue in the next section that Bose condensation can happen also far from equilibrium where typical occupancies are still large, which implies a tremendous speed-up for the condensation phenomenon as compared to the conventional thermal equilibrium situation.

\section{Bose-Einstein condensation out of equilibrium}\label{out}

We have already seen that we likely have a nonequilibrium situation for a long time. The question occurs 
whether the important features of Bose-Einstein condensation cannot 
be established much earlier than thermal equilibrium is achieved. 
This problem has been studied in~\cite{Berges:2012us,Micha:2004bv,Nowak:2013juc,BS} with a positive outcome:\footnote{Indications for similar scaling behavior have been observed also from kinetic descriptions as in Ref. \cite{Semikoz:1994zp,Semikoz:1995rd} but perturbative kinetic theory is not suitable to describe the Bose condensation dynamics in the infrared. For details about the non-perturbative lattice simulation method we consider here and limitations of perturbative kinetic descriptions, see Refs.~\cite{Berges:2010ez} and \cite{Berges:2012us}.} {\emph{Bose-Einstein condensation can happen in a non-thermal state and is established relatively fast.}}
Here we review the findings following Ref.~\cite{Berges:2012us}, where non-thermal Bose condensation has been demonstrated to occur for the relativistic as well as non-relativistic case, which we then apply and extend to the case at hand.
In the following we neglect the expansion of the Universe and just consider the relevant timescales as compared to the expansion rate.

Already shortly after the ALP field starts to oscillate the typical momenta Eq.~\eqref{momentum} are already significantly smaller than the mass, $\delta p\sim 1/3$.
Therefore, we are dealing with a non-relativistic situation.
For the considered scalar field theory, at the classical level the field is governed by the Gross-Pitaevskii equation,
\begin{equation}
\label{gpeq}
\left[i\partial_{t}+\frac{\nabla^2}{2m}-g|\psi|^2\right]\psi =0.
\end{equation}
In Appendix~\ref{nonrel} we review how this equation, featuring a conserved particle number, arises as the non-relativistic
limit of a real scalar field (which does not conserve particle number). Moreover, for the coupling one finds the relation
\begin{equation}
g=\frac{\lambda}{32 m^2}.
\end{equation}

Depending on the initial conditions, the above classical field equation can receive important corrections due to quantum-statistical fluctuations. For the considered highly occupied systems, where classical-statistical fluctuations dominate over quantum fluctuations, the dynamics can be accurately described by a classical ensemble solving the classical field equation with Monte Carlo sampling of initial conditions~\cite{Berges:2012us}. The classical-statistical nature of the dynamics has the important consequence that if we measure time in units of $t_{0}=(2m)/\Qs^{2}$ 
and momentum in units of $\Qs$ the only relevant combination of coupling and occupation number is indeed
\begin{equation}
\label{occres}
\dd\left(\frac{p}{\Qs}\right)=2mg\Qs\nocc(p)=\frac{\lambda \Qs}{16 m} \nocc(p) \, .
\end{equation}
This is shown in Appendix~\ref{apprescaled}. 

For the distribution of the classical ensemble average of modes with spatial momentum $p$,  
\begin{equation}
\nocc(t,p) \sim \langle \psi^\dagger(t,p) \psi(t,p) \rangle \, ,
\end{equation}
with conserved
\begin{equation}
N_{\rm total}=\int \frac{d^{3}p}{(2\pi)^3} \nocc(p) \, ,
\end{equation}
we use initially the same simple situation already employed to obtain the estimate~\eqref{occ}.
All modes below 
\begin{equation}
\label{qs}
\Qs=\delta p(t)=H_{1}\,\frac{a_{1}}{a(t)}\,=\frac{1}{3}m_{1}\,\frac{a_{1}}{a(t)}\,
\end{equation} 
are equally populated with the occupation number~\eqref{occ}. We start at a time $\tini\gtrsim t_{1}$ and neglect the expansion from then on.
The initial distribution is, therefore, a $\Theta$-function. 
Correspondingly the re-scaled quantity~\eqref{occres} at initial time is 
\begin{equation}
\ddzero\Theta \, \left(1-\frac{p}{\Qs}\right)\sim 20\,\kappa\,\noccfudge \,\frac{a_{1}}{\aini}\,\Theta\left(1-\frac{p}{\Qs}\right) \, .
\end{equation}
where we define,
\begin{equation}
\ddzero= \frac{\Qs}{16m}\lambda\, \nocczero,
\end{equation}
and we have inserted $m_{1}\sim m\sim const.$ and $\phi_{1}\sim\fax$ into Eqs.~\eqref{nocctyp},~\eqref{couptyp} and \eqref{qs}.

\begin{figure}[t]
    \begin{center}
    \subfigure[]{
         {\includegraphics[width=7.7cm]{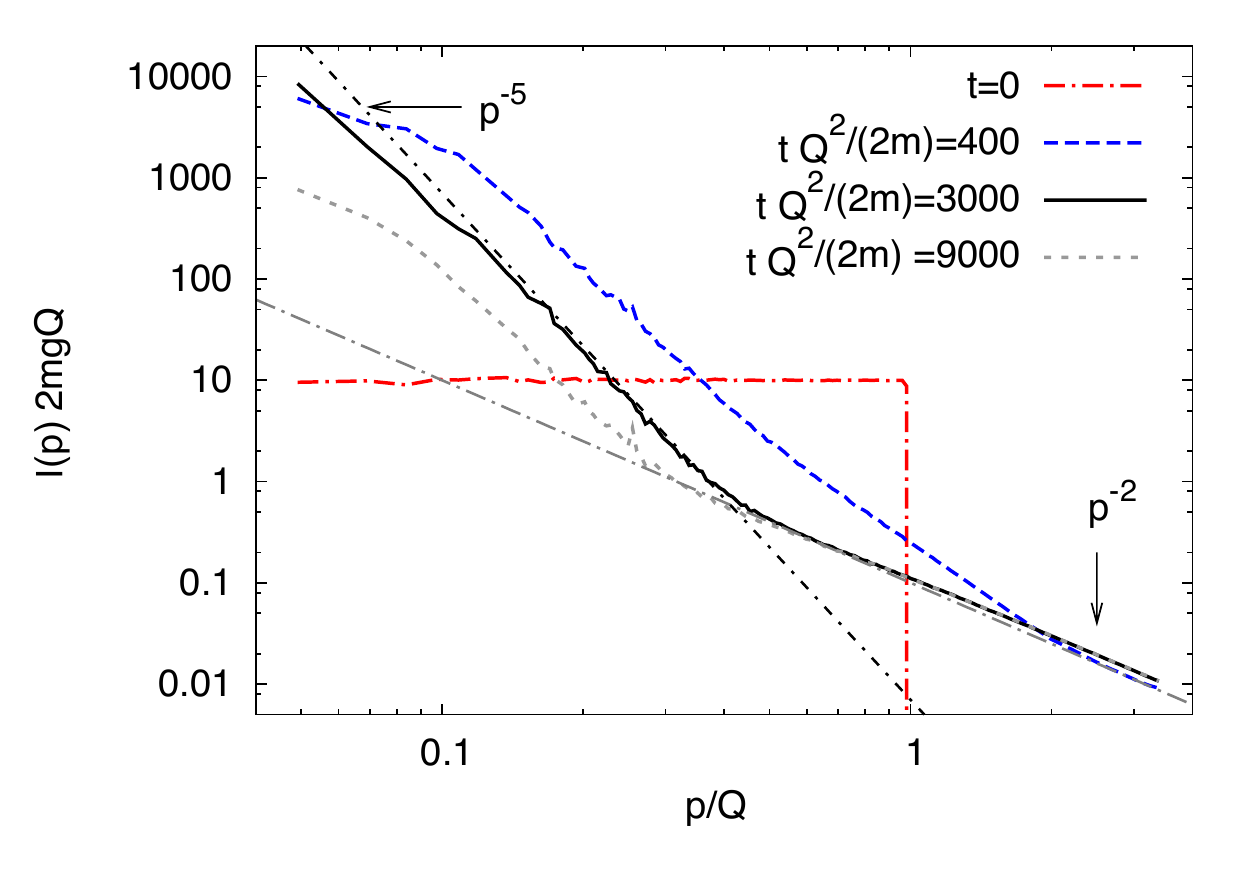}}\label{recyclefiga}}
         \hspace*{0.1cm}
              \subfigure[]{{\includegraphics[width=7.7cm,clip=true, trim= 0cm 7.5cm 0cm 5cm]{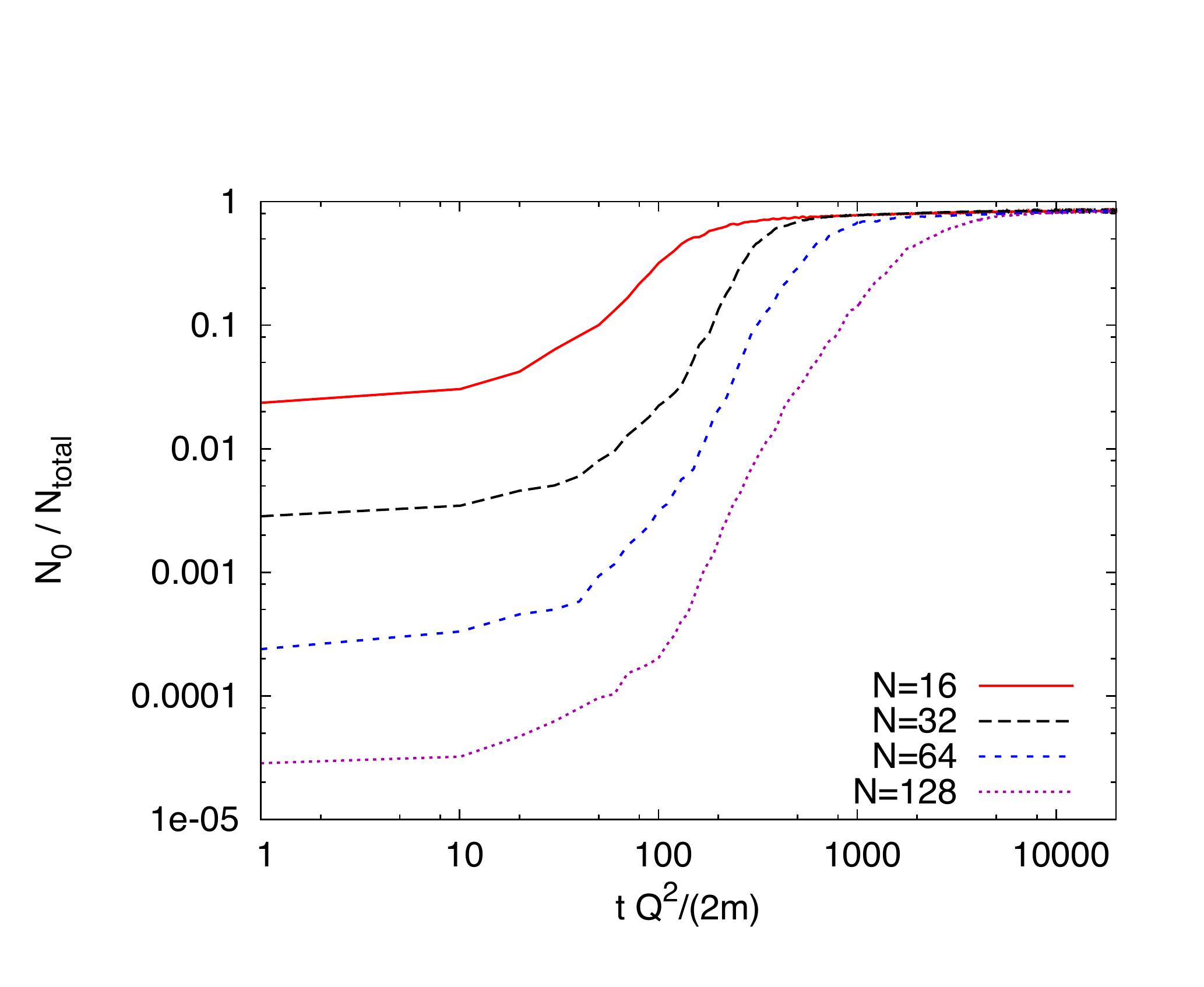}}\label{recyclefigb}}
        \caption{Time evolution of the distribution function (a) and the condensate fraction (b).}
\label{recyclefig}
    \end{center}
\end{figure}

The time evolution and the condensate formation for this type of system have been studied in Ref.~\cite{Berges:2012us}.
Fig.~\ref{recyclefiga} shows the time evolution of the distribution at different times. The $\Theta$-function shaped initial condition is seen to evolve towards a non-thermal distribution at intermediate times, characterized by a strongly enhanced power-law behavior at lower momenta. Most relevant for the question we are interested in, Fig.~\ref{recyclefigb} shows the fraction of condensed particles for an initial $\ddzero=10$ and various volumes. Here $V=16^3 \Qs^{-3}$ could be achieved by using $\aini/a_{1}\sim 10$. For $\ddzero=10$ this 
requires $\kappa\,\noccfudge\sim 5$. 

From this numerical example a couple of relevant findings can be extracted. Using Eq.~\eqref{qs} we see that,
\begin{equation}
t_{0}=\frac{18}{H_{1}}\left(\frac{\aini}{a_{1}}\right)^{2}\sim \frac{18}{\Hini}.
\end{equation}
Therefore, already one unit in our dimensionless time is an order of magnitude bigger than a Hubble time.

From Fig.~\ref{recyclefiga} we can see that even after $400t_{0}\sim 7\times 10^3/\Hini$ the distribution is not a thermal equilibrium
distribution. It deviates from the classical $1/p^2$ behavior both at low and at high momenta at this time. So the evolution is on timescales three orders of magnitude longer than the naive estimate
from the scattering between highly occupied particles.
On the other hand, as we can see from Fig.~\ref{recyclefigb}, 
\begin{equation}
\tcond\sim 400\, t_{0}\sim \frac{7\times 10^3}{\Hini}
\end{equation}
is enough to establish the condensate in rather large volumes. (See Sect.~\ref{secvolume} for a discussion of the volume dependence.)   

Importantly, this can now be compared to the timescale for thermalization as estimated in Eq.~\eqref{ttherm}, 
\begin{equation}
\ttherm\sim \frac{1}{Q} \left(\frac{1}{\lambda}\right)^\frac{1}{|\alpha|}
\sim \frac{1}{H_{1}}\frac{\aini}{a_{1}}10^{54/|\alpha|}\left(\frac{m_{1}}{\mu{\rm eV}}\right)^\frac{2}{|\alpha|}\left(\frac{10^{12}\,{\rm GeV}}{\fax}\right)^\frac{2}{|\alpha|}\gg \tcond \, .
\end{equation}
Therefore the condensate clearly forms when the system is still extremely far away from equilibrium.

Quantitatively we note that the timescale for condensate formation is somewhat larger than the naive expectation.
Therefore, for the case of a simple ALP or axion the formation turns from marginally possible within a Hubble time to being too slow.
Turned around, scalar self-interactions are therefore only then sufficient for BEC formation if either the coupling or the initial occupation number is significantly enhanced compared to the naive ALP expectation. However, as we argue in Appendix~\ref{nonrel} the former is in conflict with the non-relativistic approximation we have used and, moreover, the cosmological evolution during inflation in this case automatically pushes us to smaller field values and $\ddzero$ cannot be significantly increased in that way. Therefore the only viable option would be to increase the occupation number by decreasing the typical momentum scale compared to the naive expectation.

\section{Strongly correlated nature of the non-thermal condensate}
We have seen in the previous sections that condensate formation takes place on a fairly short timescale, where the occupation numbers are still quite far from thermal equilibrium. Therefore, the magnitude of the non-equilibrium condensate can in principle be different from that in thermal equilibrium which is established after a much longer timescale $\ttherm$.

In the following we want to emphasize that the fraction of particles in the non-equilibrium condensate established after the relatively short time $\tcond$ is not well described by a weakly interacting gas approximated by free field theory (even if it were close to thermal equilibrium).

To do this we can compare the amount of condensate formed in the non-equilibrium state and that in the thermal equilibrium of a free theory. In particular we want to compare the dependence of both condensates on the size of the coupling and the initial occupation numbers.

As discussed in the previous section and in Appendix~\ref{apprescaled}, for a given shape of the initial distribution all results, including the fraction of condensed particles,
\begin{equation}
x_{\rm cond}=\frac{N_{\rm 0}}{N_{\rm total}}
\end{equation}
in the non-equilibrium condensate depend only on $\ddzero$.
For example, keeping $\Qs$ (and $m$) fixed the non-equilibrium condensate does not change when we vary $\lambda$ but use $\nocczero\sim 1/\lambda$ such that $\nocczero\lambda$ is fixed as well.

We will now show that in the thermal equilibrium of free field theory the condensate can change under such variations in the coupling and occupation number.

In the mean-field approximation the only effect of the interaction term in the Gross-Pitaevskii equation Eq.~\eqref{gpeq} is a constant energy shift for each particle,
\begin{equation}
\Delta E =g \langle |\psi|^2\rangle=g\,\, \ndens = g \int \frac{d^{3} p}{(2\pi)^3} \nocc(p).
\end{equation}
This energy shift can simply be absorbed in a re-definition of the chemical potential.

However, we note that in the non-perturbative regime of interest for our non-equilibrium situation
the shift in energy compared to the typical relevant kinetic energies $E_{\rm typ}\sim \Qs^2/(2m)$ is not small (and cannot be made small),
\begin{equation}
\frac{\Delta E}{E_{\rm typ}}\sim \frac{\Delta E}{\Qs^2/(2m)}\sim \ddzero.
\end{equation}
This already hints at a strongly correlated system.

We now compare this to condensation in free field theory.
In equilibrium we have the condensate fraction
\begin{equation}
\label{condfraceq}
x^{\rm eq}_{\rm cond}=1-\left(\frac{T}{T_{c}}\right)^{\frac{3}{2}},
\end{equation}
with the condensation temperature
\begin{equation}
\label{tcond}
T_{c}=\frac{2\pi}{[\zeta(3/2)]^{2/3}}\frac{1}{m\,\ndens}\sim \nocczero^{2/3}.
\end{equation}

We then can determine the equilibrium temperature from energy conservation,
\begin{equation}
\label{aveenerg}
\frac{E_{\rm tot}}{N_{\rm tot}}=\frac{3\zeta(5/2)}{2\zeta(3/2)} T \left(\frac{T}{T_{c}}\right)^{\frac{3}{2}}
= \frac{\int \frac{d^{3}p}{(2\pi)^{3}} \frac{p^{2}}{2m}\nocc (p)}{\int \frac{d^{3}p}{(2\pi)^{3}}\nocc (p)}=\frac{3 \Qs^{2}}{10\, m}.
\end{equation}

Combining Eqs.~\eqref{tcond} and Eq.~\eqref{aveenerg} we find that the condensed fraction is
\begin{equation}
1-\left(\frac{T}{T_{c}}\right)^{\frac{3}{2}}\approx 1- 0.25 \nocczero^{-\frac{2}{5}},
\end{equation}
independent of the coupling $\lambda$ (as it should be in equilibrium).
Therefore in the free field equilibrium the condensate fraction~\eqref{condfraceq}  {\emph{does depend}} on $\nocczero$ even if we keep $\lambda\nocczero$ fixed. This shows that the two condensates are quantitatively not the same.

We note that since in the situation of interest we expect a fairly high condensate fraction, a better quantity to highlight this difference is actually the ratio of the fractions of non-condensed particles $1-x_{\rm cond}$ in and out of equilibrium.

\section{Volume dependence of the condensation timescale}\label{secvolume}
An additional important feature of the dynamics is that the timescale for condensation is volume dependent.

\begin{figure}[t]
    \begin{center}
         {\includegraphics[scale=0.60]{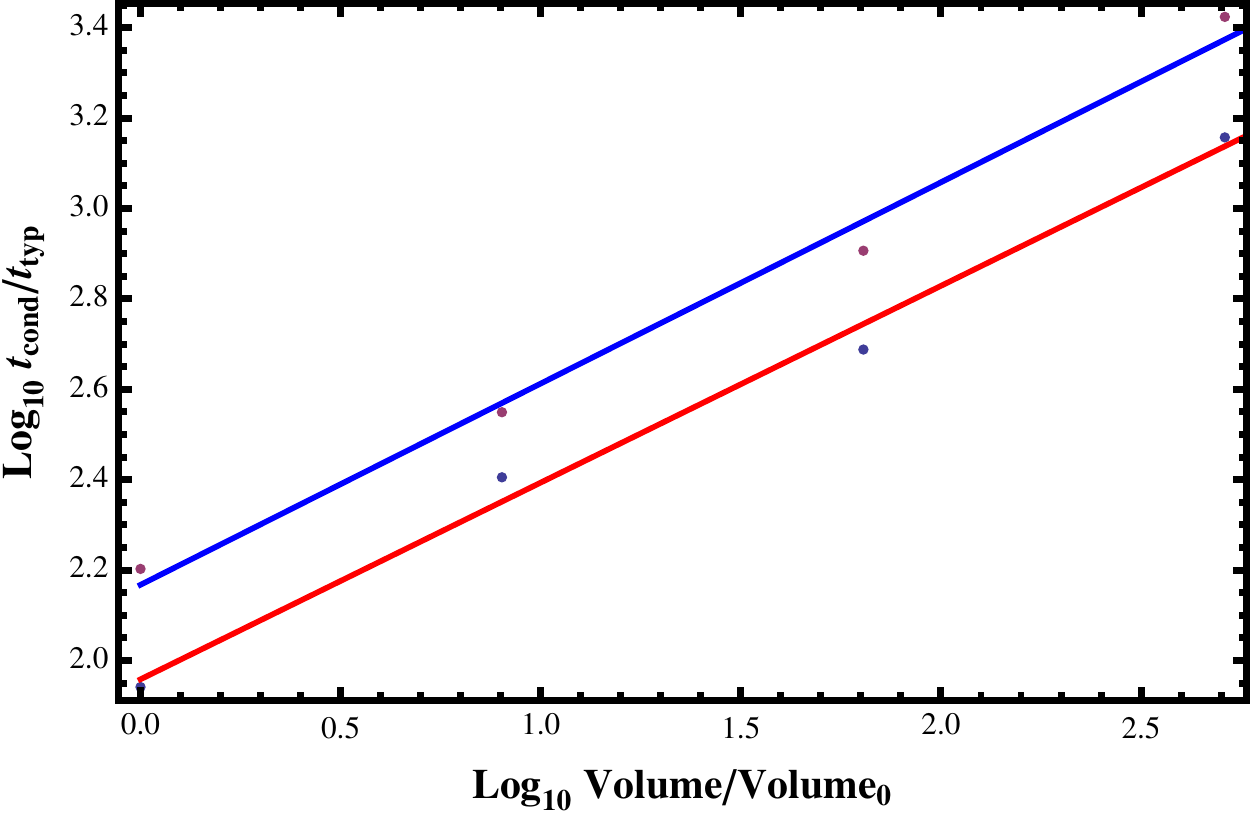}}
         \caption{Dependence of the condensate formation time on the volume of the patch. The red curve is for a condensate fraction of 60\% and the blue curve is for 30\%. Fitting this suggests a behavior
         $\sim V^{0.3-0.55}\sim L^{0.9-1.7}$.}
         \label{timedepfig}
    \end{center}
\end{figure}

This can be directly observed in Fig.~\ref{recyclefigb} where one can see that in order to achieve a given fixed level of condensate, i.e.~a fixed value of $N_{0}/N_{\rm total}$ in the picture, more time is needed when the volume is larger. 
This is not too surprising. Naively one first expects local patches of condensate to form which then grow over time.
This  is consistent with the results of Fig.~\ref{timedepfig} where the time required for the condensate to reach a condensate fraction of 30\% (blue) and 60\% (red) is shown. We can see that there is a sizeable volume dependence.

Above we have asked for the time required to establish a certain fraction of condensate in a given volume. 
One could also ask a slightly different question. Is there a typical patch size on which condensation occurs first (similar to magnetic domains in a magnet). The later volume dependence of condensation in larger volumes then arises from the time required to grow/merge these patches.
While our analysis is not sufficient to fully answer this question an indication can be found in Fig.~\ref{recyclefigb}, where the rise in the condensate fraction starts roughly at the same time $\sim 10 \,t_{0}$ independent of the volume.

For axion-like particle dark matter both effects could be relevant depending on the question one asks, since very different length scales may be relevant. For effects in structure formation one would require condensation on galactic scales, whereas in experiments for the direct detection of axions already condensates on much smaller length scales may be noticeable.

\section{Conclusions}
We have analyzed the question of Bose-Einstein condensation for axion-like particles (ALPs) via their self-couplings.
Our main findings are
\begin{itemize}
\item{} Thermalization, i.e. establishing a true thermal distribution with typical occupation numbers of order one for $p^{2}/(2m) \sim T$, is  slow for weak coupling, even if
we have extremely high initial occupation numbers. 
\item{} The formation time for a condensate $\tcond$ can be much smaller than the time required for thermalization $\ttherm$. The corresponding timescale is similar (albeit a bit longer) than that obtained from an estimate of the Bose-enhanced interaction rates.
\item{} Condensate formation takes place in a non-equilibrium state and the condensate formed in this state is parametrically different from that obtained for a weakly interacting gas in thermal equilibrium.
\end{itemize}

In this note we have only considered scalar self-interactions. As already pointed out in~\cite{Sikivie:2009qn} these interactions are not sufficient
when it comes to a condensate formation in the expanding early Universe. 
Yet, importantly it can serve as a simple model for the basic dynamics of the formation of a Bose-Einstein condensate of ALP dark matter by Bose-enhanced interactions. 
Our positive finding that condensate formation takes indeed place on the corresponding short timescale now motivates studying the phenomenologically more relevant interactions via gravity.

\section*{Acknowledgements}
We would like to thank T.~Gasenzer and D.~Sexty for very helpful discussions and collaborations on related work.

\begin{appendix}
\section{The non-relativistic limit of a real scalar field}\label{nonrel}
\subsection{From the real, nonlinear Klein-Gordon equation to the Gross-Pitaeevski equation}
The starting point for axions is a relativistic theory for a real scalar field. Importantly there is no conserved particle number.
However, at the very small momenta we are interested in, i.e. in the non-relativistic regime, these are just spinless particles whose number is conserved. 
In this appendix let us briefly sketch how the non-relativistic equation with its conserved particle number arises from the full relativistic theory.

The Lagrangian in the relativistic case is,
\begin{equation}
{\mathcal L}_{\rm rel}=\frac{1}{2}(\partial_{\mu}\phi)^{2}-\frac{m^{2}}{2}\phi^2-\frac{\lambda}{4!}\phi^4.
\end{equation}
This gives the equation of motion
\begin{equation}
\label{eomrel}
\Box \phi+m^{2}\phi +\frac{\lambda}{6}\phi^3=0.
\end{equation}

To obtain the non-relativistic limit we can now factor out the rapid oscillation caused by the mass in 
\begin{equation}
E\approx m+\frac{p^2}{2m}+\ldots.
\end{equation}
This can be done with the ansatz,
\begin{equation}
\phi=\Re[\exp(-imt)\tilde{\psi}],
\end{equation}
where we have taken the real part, because $\phi$ is a real field.

For this to be a valid approximation actually requires two conditions to be fulfilled. One is the obvious $p\ll m$.
However, it also requires that the mass itself is time independent, i.e. that the contribution of self-interactions to the mass 
$\sim (\lambda/6)\phi^2\ll m^2$. We will comment on the latter below.

Having factored out the fast oscillation $\tilde{\psi}$ is a slowly changing field with 
\begin{equation}
\left|\frac{\partial^2\tilde{\psi}}{\partial t^{2}}\right|\ll \left|m\frac{\partial\tilde{\psi}}{\partial t}\right|.
\end{equation}
Neglecting these terms and inserting into Eq.~\eqref{eomrel} we obtain,
\begin{equation}
\label{inter}
\Re[(-2im\frac{\partial\tilde{\psi}}{\partial t}-\nabla^2\tilde{\psi})\exp(-imt)]+\frac{\lambda}{6}(Re[\tilde{\psi}\exp(-imt)])^3=0.
\end{equation}

Since we are after the slowly varying parts only, we can remove all parts that oscillate with a frequency much larger than $m$, i.e.
in particular we neglect all parts that change with frequency $3m$,
\begin{equation}
\label{inter2}
\Re\left[\left\{-2im \frac{\partial\tilde{\psi}}{\partial t}-\nabla^2\tilde{\psi}+\frac{\lambda}{8}\tilde{\psi}|\tilde{\psi}|^2\right\}\exp(-imt)\right]=0.
\end{equation}
To good approximation we can imagine that the faster oscillating parts average to zero in the evolution of $\tilde{\psi}$.
Physically neglecting these fast oscillating terms corresponds to eliminating the particle number changing processes that are 
energetically forbidden for slow, non-relativistic particles.

Solving the full complex equation
\begin{equation}
-2im\frac{\partial \tilde{\psi}}{\partial t}-\nabla^2\tilde{\psi}+\frac{\lambda}{8}\tilde{\psi}|\tilde{\psi}|^2=0
\end{equation}
automatically gives a solution of Eq.~\eqref{inter2}.

Finally, inserting
\begin{equation}
\tilde{\psi}\rightarrow\frac{\psi}{\sqrt{2m}}
\end{equation}
we find
\begin{equation}
i\frac{\partial\psi}{\partial t}=-\frac{\nabla^2}{2m}\psi+\frac{\lambda}{32 m^2}\psi|\psi|^2,
\end{equation}
which has the form of a non-linear Schroedinger equation also called Gross-Pitaevskii equation.
Comparison with the usual form gives the coupling
\begin{equation}
g=\frac{\lambda}{32m^2}.
\end{equation}

\subsection{Limit of applicability of the non-relativistic equation for high densities}
As already mentioned above, validity of the non-relativistic limit also poses constraints on the attainable field values and therefore the occupation numbers.
Let us briefly consider what this implies for the quantity 
\begin{equation}
\label{dimless}
\ddzero=2mg\Qs \nocczero=\frac{\Qs}{16m}\lambda \nocczero
\end{equation}
that is the relevant quantity for the evolution in the classical statistical regime 
(see Sect.~\ref{out} and App.~\ref{apprescaled}).

As we mentioned above, the non-relativistic limit requires
\begin{equation}
\phi\lesssim \frac{m}{\sqrt{\lambda}}.
\end{equation}
Then the total particle number density is
\begin{equation}
\ns\sim \frac{V(\phi)}{m}\lesssim \frac{m^{3}}{\lambda}.
\end{equation}
Using our $\Theta$-function shaped momentum space distribution from Eq.~\eqref{thetainitial} we have
\begin{equation}
\ns\sim \nocczero\Qs^3.
\end{equation}
Inserting into Eq.~\eqref{dimless} we find,
\begin{equation}
\ddzero\lesssim \frac{m^2}{\Qs^2},
\end{equation}
independent of $\lambda$.

Therefore to increase $\ddzero$ and therefore push deeper into the non-relativistic regime, we cannot increase the coupling, but instead we have to go to lower and lower momenta.

\subsection{Cosmological evolution limits us to non-relativistic field values}
Above we have argued that our non-relativistic approximation is limited to field values $\phi\lesssim m/\sqrt{\lambda}$ and that this in turn
limits the size of $\ddzero$ which determines the evolution of the classical field.
Interestingly the cosmological evolution automatically pushes us into this regime, thereby limiting the achievable values of $\ddzero$ unless the typical momentum scale is lowered.

To see that this is the case let us consider Eq.~\eqref{eomrel} for a spatially constant field but in an expanding universe,
\begin{equation}
\ddot{\phi}+3H\dot{\phi}+m^{2}\phi+\frac{\lambda}{6}\phi^3=0.
\end{equation}

In the region $\phi\gg m/\sqrt{\lambda}$ we can neglect the mass term.

We can now consider the evolution of the energy density in this field,
\begin{equation}
\frac{d\rho}{dt}=\frac{d}{dt}\left(\frac{1}{2}\dot{\phi}^2+\frac{\lambda}{4!}\phi^4\right)=-3H\dot{\phi}^2=-6HT_{\rm kin},
\end{equation}
where we have used the equation of motion in the second to last step.

If $H\ll \sqrt{\lambda}{\phi}$ the equation describes a weakly damped nonlinear oscillation. To see the effect of the damping we can use the virial theorem for this type of potential,
\begin{equation}
\langle T_{\rm kin}\rangle=\langle \frac{1}{2}\dot{\phi}^2\rangle=\langle\frac{4}{6}\rho\rangle,
\end{equation}
where the average is taken over an oscillation.
Using this we have,
\begin{equation}
\frac{d}{dt}\langle\rho\rangle=-4H\langle \rho\rangle.
\end{equation}
Accordingly
\begin{equation}
\langle \rho\rangle\sim 1/a^{4}(t),
\end{equation}
from which we can extract for the field amplitude
\begin{equation}
\phi(t)\sim \frac{1}{a(t)}.
\end{equation}

In contrast to the situation where we are dominated by the mass (and have an overdamped oscillator) the field amplitude decays.
In the end (for not-quite constant decreasing $H$) this damping is only stopped when the mass term becomes dominant and
\begin{equation}
\phi\sim \frac{m}{\lambda}.
\end{equation}
As already described in the above subsection this limits the maximal value of the parameter $\ddzero$.

\section{Re-scaled scalar field evolution}\label{apprescaled}
For a given shape of the distribution of the occupation numbers, such as, e.g. the $\Theta$-function distribution
assumed in Sect.~\ref{out} we can perform a re-scaling in momentum and one in the overall height of the distribution without changing the shape.
Moreover, one can wonder how the evolution changes under a change in the coupling constant.
We now determine the appropriate variables for which the time-evolution of the classical is independent of these re-scalings.

First we note that in classical evolution we can always re-scale the coupling to one.
Beginning with the Gross-Pitaevskii equation
\begin{equation}
\left[i\partial_{t}+\frac{\nabla^2}{2m}-g|\psi|^2\right]\psi=0,
\end{equation}
we can use
\begin{equation}
\psi=\frac{\psi^{\prime}}{\sqrt{g}}
\end{equation}
to obtain
\begin{equation}
\left[i\partial_{t}+\frac{\nabla^2}{2m}-|\psi^{\prime}|^{2}\right]\psi^{\prime}=0.
\end{equation}

When re-scaling our distribution in the momentum direction it is clear that the typical space/momentum scales as well
as the typical scale for the time-evolution change. We therefore introduce appropriate dimensionless variables that describe 
the time-evolution and the spatial behavior,
\begin{equation}
\tau =\frac{t}{t_{0}},\quad \varkappa=x \Qs,\quad p=\Qs p^{\prime}
\end{equation}
where $\Qs$ is the typical momentum scale of the distribution and
\begin{equation}
t_{0}=\frac{2m}{\Qs^{2}}.
\end{equation}
Doing a further re-scaling
\begin{equation}
\psi^{\prime}=\frac{\psi^{\prime\prime}}{\sqrt{t_{0}}},
\end{equation}
we arrive at the final dimensionless form,
\begin{equation}
\label{rescaled}
\left[\partial_{\tau}+\nabla^{2}_{\varkappa}-|\psi^{\prime\prime}|^2\right]\psi^{\prime\prime}=0.
\end{equation}

The only part of Eq.~\eqref{rescaled} affected by the re-scaling is now  the term $|\psi^{\prime\prime}|^2$.
The final step is therefore to relate this to the re-scaling of the distribution of the occupation numbers.
For this we have to un-do the effects of the field re-scalings as well as take into account
that the total number of occupied modes, i.e. the total number of particles and therefore $|\psi^{\prime\prime}|^2$
is changed by a volume factor $Q^3$ in momentum space,
\begin{equation}
N_{\rm total}=\int \frac{d^{3}p}{(2\pi)^3} \nocc(p)= Q^{3}\int \frac{dp^{\prime\,3}}{(2\pi)^{3}} \hat{\nocc}(p^{\prime}),
\end{equation}
where $\hat{\nocc}(p^{\prime})=\nocc(\Qs p^{\prime})$ gives the shape of the distribution and is unaffected by changes of $\Qs$.
Combining this we have
\begin{equation}
|\psi^{\prime\prime}|^2=t_{0}g |\psi|^2\sim t_{0} g Q^{3} \hat{\nocc}\sim 2mg\Qs \hat{\nocc}.
\end{equation}
Therefore under the discussed re-scalings the relevant quantity is,
\begin{equation}
\dd\left(\frac{p}{\Qs}\right)= 2mg\Qs\nocc(\Qs p^{\prime})= \frac{\lambda \Qs}{16m}\nocc(\Qs p^{\prime}).
\end{equation} 
which is what is shown in the figures.
\end{appendix}

\bibliographystyle{h-physrev5}
\bibliography{revised.bbl}

\begin{thebibliography}{10}

\bibitem{Preskill:1982cy}
J.~Preskill, M.~B. Wise, and F.~Wilczek,
\newblock Phys. Lett. {\bf B120}, 127 (1983).

\bibitem{Abbott:1982af}
L.~F. Abbott and P.~Sikivie,
\newblock Phys. Lett. {\bf B120}, 133 (1983).

\bibitem{Dine:1982ah}
M.~Dine and W.~Fischler,
\newblock Phys. Lett. {\bf B120}, 137 (1983).

\bibitem{Sikivie:2009qn}
P.~Sikivie and Q.~Yang,
\newblock Phys.Rev.Lett. {\bf 103}, 111301 (2009), arXiv:0901.1106.

\bibitem{Sikivie:2012gi}
P.~Sikivie,
\newblock (2012), arXiv:1210.0040.

\bibitem{Banik:2013rxa}
N.~Banik and P.~Sikivie,
\newblock Phys.Rev. {\bf D88}, 123517 (2013), arXiv:1307.3547.

\bibitem{Duffy:2006aa}
L.~D. Duffy {\em et~al.},
\newblock Phys. Rev. {\bf D74}, 012006 (2006), arXiv:astro-ph/0603108.

\bibitem{Irastorza:2011gs}
I.~G. Irastorza {\em et~al.},
\newblock JCAP {\bf 1106}, 013 (2011), arXiv:1103.5334.

\bibitem{Jaeckel:2013sqa}
J.~Jaeckel and J.~Redondo,
\newblock JCAP {\bf 1311}, 016 (2013), arXiv:1307.7181.

\bibitem{Erken:2011vv}
O.~Erken, P.~Sikivie, H.~Tam, and Q.~Yang,
\newblock Phys.Rev.Lett. {\bf 108}, 061304 (2012), arXiv:1104.4507.

\bibitem{Erken:2012dz}
O.~Erken, P.~Sikivie, H.~Tam, and Q.~Yang,
\newblock Phys.Rev. {\bf D85}, 063520 (2012), arXiv:1111.1157.

\bibitem{Saikawa:2012uk}
K.~Saikawa and M.~Yamaguchi,
\newblock Phys.Rev. {\bf D87}, 085010 (2013), arXiv:1210.7080.

\bibitem{Davidson:2013aba}
S.~Davidson and M.~Elmer,
\newblock JCAP {\bf 1312}, 034 (2013), arXiv:1307.8024.

\bibitem{Noumi:2013zga}
T.~Noumi, K.~Saikawa, R.~Sato, and M.~Yamaguchi,
\newblock Phys.Rev. {\bf D89}, 065012 (2014), arXiv:1310.0167.

\bibitem{Berges:2012us}
J.~Berges and D.~Sexty,
\newblock Phys.Rev.Lett. {\bf 108}, 161601 (2012), arXiv:1201.0687.

\bibitem{Micha:2004bv}
R.~Micha and I.~I. Tkachev,
\newblock Phys.Rev. {\bf D70}, 043538 (2004), arXiv:hep-ph/0403101.

\bibitem{Nowak:2013juc}
B.~Nowak {\em et~al.},
\newblock (2013), arXiv:1302.1448.

\bibitem{BS}
N.~G. Berloff and B.~V. Svistunov,
\newblock Phys. Rev. A {\bf 66}, 013603 (2002).

\bibitem{Arias:2012ly}
P.~Arias {\em et~al.},
\newblock JCAP {\bf 1206}, 013 (2012), arXiv:1201.5902.

\bibitem{Jaeckel:2010ni}
J.~Jaeckel and A.~Ringwald,
\newblock Ann. Rev. Nucl. Part. Sci. {\bf 60}, 405 (2010), arXiv:1002.0329.

\bibitem{Sikivie:2006ni}
P.~Sikivie,
\newblock Lect.Notes Phys. {\bf 741}, 19 (2008), arXiv:astro-ph/0610440.

\bibitem{Berges:2013lsa}
J.~Berges, K.~Boguslavski, S.~Schlichting, and R.~Venugopalan,
\newblock JHEP {\bf 1405}, 054 (2014), arXiv:1312.5216.

\bibitem{Berges:2008sr}
J.~Berges and G.~Hoffmeister,
\newblock Nucl.Phys. {\bf B813}, 383 (2009), arXiv:0809.5208.

\bibitem{Semikoz:1994zp}
D.~Semikoz and I.~Tkachev,
\newblock Phys.Rev.Lett. {\bf 74}, 3093 (1995), arXiv:hep-ph/9409202.

\bibitem{Semikoz:1995rd}
D.~Semikoz and I.~Tkachev,
\newblock Phys.Rev. {\bf D55}, 489 (1997), arXiv:hep-ph/9507306.

\bibitem{Berges:2010ez}
J.~Berges and D.~Sexty,
\newblock Phys.Rev. {\bf D83}, 085004 (2011), arXiv:1012.5944.

\end{thebibliography}
\end{document}